\documentstyle[aas2pp4]{article}
 
\begin{document}
 
\twocolumn

\title{Discovery of seven T Tauri stars and a brown dwarf candidate in
the nearby TW Hydrae Association.}

\author{R. A. Webb\altaffilmark{1}, B. Zuckerman\altaffilmark{1},
I. Platais\altaffilmark{2}, J. Patience\altaffilmark{1},
R. J. White\altaffilmark{1}, M. J. Schwartz\altaffilmark{1},
C. McCarthy\altaffilmark{1}}

\altaffiltext{1}{UCLA Division of Astronomy and Astrophysics;
Los Angeles, CA 90095-1562; webbr@astro.ucla.edu}

\altaffiltext{2}{Dept. of Astronomy; Yale University; P.O. Box 208101; New
Haven, CT 06520-8101}

\begin{abstract}

       We report the discovery of five T Tauri star systems, two of
which are resolved binaries, in the vicinity of the nearest known
region of recent star formation, the TW Hydrae Association.  The newly
discovered systems display the same signatures of youth (namely high
X-ray flux, large Li abundance and strong chromospheric activity) and
the same proper motion as the original five members.  These
similarities firmly establish the group as a bona fide T Tauri
association, unique in its proximity to Earth and its complete
isolation from any known molecular clouds.

        At an age of $\sim$10 Myr and a distance of $\sim$50 pc, the
association members are excellent candidates for future studies of
circumstellar disk dissipation and the formation of brown dwarfs and
planets.  Indeed, as an example, our speckle imaging revealed a faint,
very likely companion 2\arcsec~north of CoD-33$^{\circ}$ 7795 (TWA 5).
Its color and brightness suggest a spectral type $\sim$M8.5 which, at
an age of $\sim$10$^7$ years, implies a mass $\sim$ 20 M$_{\rm
Jupiter}$.  

\end{abstract}

\keywords{open clusters and associations: individual (TW Hya) ---
stars: low-mass, brown dwarfs --- stars: pre-main sequence --- X-rays:
stars}

\section{Introduction}

    The origin of the T Tauri star TW Hya has long been a mystery to
astronomers as it is far from any known progenitor cloud.  Excepting
its isolation, Rucinski \& Krautter (1983) noted that TW Hya has all
the characteristics of a "classical" T Tauri star.  Subsequent
searches by de la Reza et al. (1989) and Gregorio-Hetem et al. (1992) at
IRAS Point-Source Catalog positions revealed four other T Tauri
systems in the same region of the sky and suggested the stars might be
associated.  Strong X-ray fluxes seen at all five of these systems led
Kastner et al. (1997) to conclude that the group does indeed form a
physical association (which they dubbed the "TW Hya Association",
hereafter TWA), at a distance of only $\sim$ 50 pc and an age $\sim$
20 Myr.  It is remarkable that the closest known region of recent star
formation is only now being identified and cataloged.

        The strength of the X-ray emission tabulated by Kastner et
al. (1997) suggested that we might find additional TWA members at
positions of X-ray bright stars.  We selected sources from the ROSAT
All-Sky Survey within 12$^{\circ}$ of the approximate cluster center
(11$\rm{^h}$ 15$\rm{^m}$, -33$^{\circ}$) and cross-correlated their
positions with stars in the Hubble Guide Star \& USNO A1.0 catalogs.
Since this search region passed close to another young, nearby star,
HR 4796 (Jura et al. 1998; Stauffer et al. 1995), we enlarged our
search area to include an overlapping region of radius 8$^{\circ}$
centered on HR 4796 (12$\rm{^h}$ 35$\rm{^m}$, -40$^{\circ}$).
Throughout this paper we consider HR 4796 A \& B as TWA members (see
section 3).  Our candidates were prioritized using X-ray flux, X-ray
hardness ratios and proper motion.

        Results discussed below indicate that our search criteria were
appropriate, we discovered five T Tauri systems containing a total
of seven stars; all are likely TWA members.  In addition, we found a
$\sim$20 Jupiter mass probable companion to CoD -33$^{\circ}$7795
(TWA 5), a previously known TWA member.

\section{Observations}

        We used the Southern Proper Motion (SPM) program plates
(Platais et al. 1995; Girard et al. 1998) to derive proper motions of
potential TWA members.  The SPM plates contain multiple images of
bright stars in the TWA region, with a mean epoch difference ranging
from 20 to 25 years.  Several of the brighter, previously known TWA
members were measured and reduced using the standard SPM reduction
procedure (e.g. Girard et al. 1998) which gives absolute proper
motions accurate to about 2-5 mas yr$^{-1}$ depending upon the star's
magnitude. For the remainder of candidates we used a simplified plate
calibration to derive the proper motions with an accuracy of about 5-8
mas yr$^{-1}$.

        On 1998 Feb. 6-7 (UT) we obtained spectra from
$\sim$6350\AA~to 7735\AA~of approximately 50 candidate stars with the
LRIS spectrometer (Oke et al. 1995) on the W. M. Keck II telescope.
The 0\farcs 7 slit yielded a measured resolution of $\sim$1.8\AA.
Spectra were extracted and calibrated using IRAF.  Seven stars,
including the individual components of two double systems, show
H$\alpha$ emission and strong Li absorption, characteristics of T
Tauri stars.  We also obtained LRIS I-band images of the fields
surrounding several TWA stars.

        Speckle observations at the NASA IRTF 3m telescope were
conducted 1998 Feb. 18-20 (UT) at 2.2 $\mu$m to search for close
companions to the newly discovered systems as well as four of the
original TWA members and HR 4796B.  These diffraction-limited
observations resolved the known binary stars Hen 600 (Sep = $1\farcs
44 \pm 0\farcs 01$, PA = $215.4^{\circ} \pm1.0^{\circ}$) and CoD
-29$^{\circ}$ 8887 (Sep = $0\farcs 56 \pm 0\farcs 01$, PA =
$29.5^{\circ} \pm 1.0^{\circ}$).  One new object was discovered, a
faint companion to TWA 5 (see Discussion).  Otherwise these
observations preclude the presence of any unseen companions down to a
typical $\Delta$K = 4 magnitudes over the separation range from
0\farcs 2 to 1\farcs 0 as derived by determining the maximum
amplitudes of cosine waves that could be hidden in the noise of the
power spectra (cf. Ghez et al. 1993, Henry 1991).  To search for wider
companions ($\sim$ 1\arcsec-3\arcsec), individual images were aligned
on the brightest speckle and summed; these shift-and-add data are
sensitive to a limiting $\Delta$K $\sim$ 6 magnitudes.  

\section{Results and Discussion}

        LRIS spectra of the 7 newly discovered young stars and 6 of
the previously known TWA members are shown in Fig. 1a and 1b.  All of
the stars show the Li I (6708\AA) line in strong absorption with
equivalent widths (EW) ranging from $\sim$ 0.36 \AA ~to 0.57 \AA
~(Table 1).  The strengths of the Li lines are typical of T Tauri
stars and distinctly above that found for Pleiades stars at $\sim$
10$^8$ years (e.g. Mart\'{i}n 1997).  Each star, except HD 98800, has
H$\alpha$ in strong emission.  Following the spectroscopic T Tauri
classification scheme of Mart\'{i}n (1998), 9 of the 13 are classified
as true T Tauri stars while the remaining 4 fall slightly (but within
the 50 m\AA~uncertainty) below the Li EW strength threshold into the
``post - T Tauri'' regime.  Of those classified as T Tauri stars, 3
have H$\alpha$ emission above the threshold (Mart\'{i}n 1997) for
``classical'' as opposed to ``weak-lined'' T Tauri stars.  This
evolutionary state agrees with a cluster age of $\sim$ 10 Myr as
discussed below.

\begin{figure}[t]
\figurenum{1}
\epsscale{1.0}
\plotone{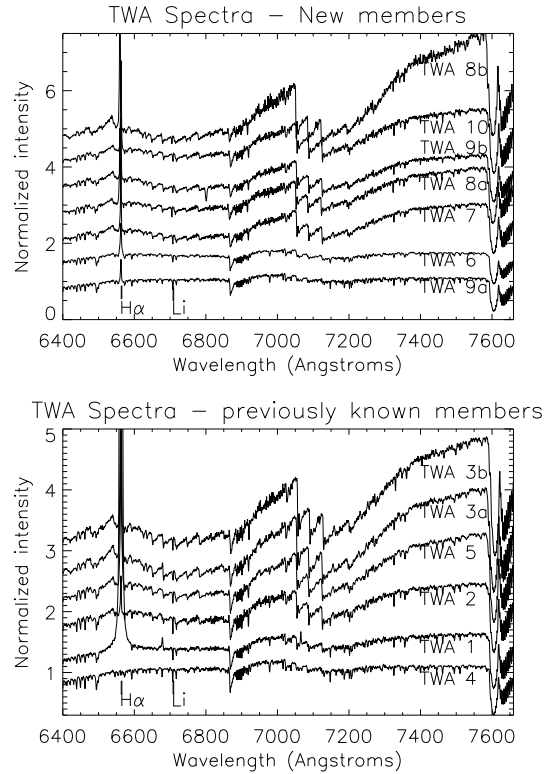}
\caption{\scriptsize LRIS spectra of a.)  the seven new and b.) six
of the original members of the TW Hya Association.  In each plot,
stars are ordered by spectral type with the earliest types at bottom.
The H$\alpha$ lines overlap and are confused, but their EWs can be
found in Table 1 as can the derived spectral types.  H$\alpha$
(emission - 6563\AA) and Li I (absorption - 6708\AA), both indicators
of youth, are prominent in the spectra.}
\end{figure}

        Each of the newly discovered systems, along with the original
five members, were strongly detected in the ROSAT All-Sky Survey.  The
X-ray luminosities normalized to the total stellar luminosities are
listed in Table 1.  All five new systems display a similarly large
L$_{\rm x}$ / L$_{\rm bol}$ consistent with Kastner et al.'s age
determination (20 +/- 10 million years) placing them at a seldom
studied evolutionary state -- the transition between the T Tauri and
the post-T Tauri phase.

     Hipparcos observed four TWA members (HD 98800, TW Hya, TWA 9, \&
HR 4796) providing accurate parallactic distances and proper motions.
The Hipparcos distances to TW Hya and HD 98800, 56 pc and 47 pc
respectively, are in excellent agreement with the estimates of Kastner
et al. (1997); TWA 9 and HR 4796 are at 50 pc and 67 pc, respectively.
Proper motions of other members were extracted from the SPM database
(see Table 1 and Fig. 2).  While there is some dispersion, the proper
motion among the TWA members is fairly uniform.  Whether or not the
motions are consistent with a common origin is controversial (Jensen
et al. 1998, Soderblom et al. 1998, Hoff et al. 1998).  We defer
discussion to a second paper in which we use echelle radial velocities
to demonstrate common space motions for TWA members (Webb, Reid \&
Zuckerman 1998).

\begin{figure}[t]
\figurenum{2}
\epsscale{1.0}
\plotfiddle{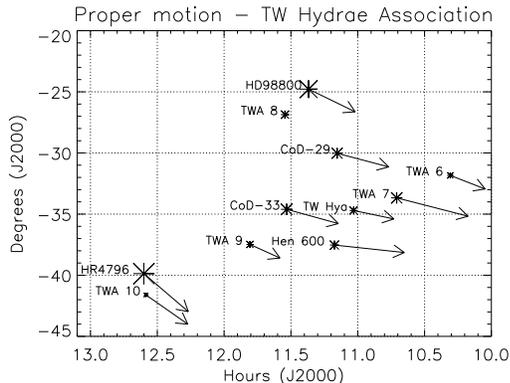}{1.75in}{90}{30}{30}{110}{-30}
\caption{\scriptsize Positions and motions of the TW Hya Association
stars.  Coordinates are equinox 2000.  Vectors show the observed
proper motion extrapolated over a period of 200,000 years.  Proper
motion has not yet been measured for TWA 8.  Solar reflex motion is
responsible for a substantial portion of the observed proper motion.
Assuming a mean distance of 50 pc for the association, a typical
correction for the Sun's motion relative to the LSR is +55 mas
yr$^{-1}$ in RA and +20 mas yr$^{-1}$ in Declination.  However, at
this distance, a substantial uncertainty is introduced when making
these corrections (Hoff et al. 1998).  The symbol size represents the
J-band flux.}
\end{figure}

    Given that we searched only in a limited solid angle of sky, one
might ask whether a comparable spectroscopic survey of X-ray bright
stars, but in a very different direction, would reveal many nearby
stars of comparable young age.  We think this is unlikely and that the
TWA represents a true localized association.  Support for this
assertion is threefold:

    First, consider the population of nearby ($<$ 70 pc) stars with T
Tauri or A star characteristics which are far from dark clouds and
which exhibit substantial excess IR emission (defined here as a ratio
of excess far-IR luminosity divided by total bolometric luminosity of
10$^{-3}$ or greater).  Many stars which appear to fulfil all these
criteria have been shown to be peculiar post - main sequence stars.
(e.g., Zuckerman et al 1995; Fekel et al 1996; Webb et al 1999 - in
prep).  Indeed, we know of only 6 stars that satisfy the preceding
criteria - four of the six are in the TWA (HD 98800, TW Hya, Hen 600,
and HR 4796).  The other two are $\beta$ Pic and 49 Cet.  This
concentration points to the vicinity of TW Hya as special.  Similarly,
Jensen et al. (1998) note an excess of young, low-velocity stars in
the vicinity of TW Hya.
	
    Second, surveys of X-ray bright stars in other regions well
separated from known dark clouds do not produce a similar number of T
Tauri stars.  In our LRIS survey, 12 of our targets were late-type
stars (spectral types K5 or later) and 5 of these have T Tauri
spectral characteristics.  We consider only late-type stars because
the Li depletion time is short and there is a clear distinction in Li
EW between T Tauri (10 Myr) and Pleiades (120 Myr) age stars.  Magnani
and colleagues have searched for T Tauri stars toward 3 widely
separated translucent clouds: MBM 40 (Magnani et al 1996) and MBM 7
and 55 (Hearty et al 1998) which are located at $\vert$b$\vert$ $\sim$
40 degrees.  Their target list consisted of X-ray bright ROSAT stars;
they list 42 with spectral type similar to the TWA members.  None
showed substantial lithium absorption in vivid contrast to results of
our LRIS survey near TW Hya.  Also, in 1998 September we surveyed 11
late K and M-type, X-ray bright field stars, many of which are listed
in Fleming (1998), with the Kast spectrograph on the 3 m Shane
telescope at Lick Observatory.  None showed any detectable Li with an
upper limit of $\sim$ 50 m\AA.

    And third, if T Tauri stars are truly widely distributed in the
solar neighborhood, then we would anticipate seeing many X-ray bright
post T Tauri stars in the 10-100 million year age range.  As we
mention below, our LRIS data show no evidence of such a population.
Together these characteristics are adequate to firmly establish the
stars in Table 1 as a physical association.

\begin{figure}[t]
\figurenum{3}
\epsscale{0.75}
\plotfiddle{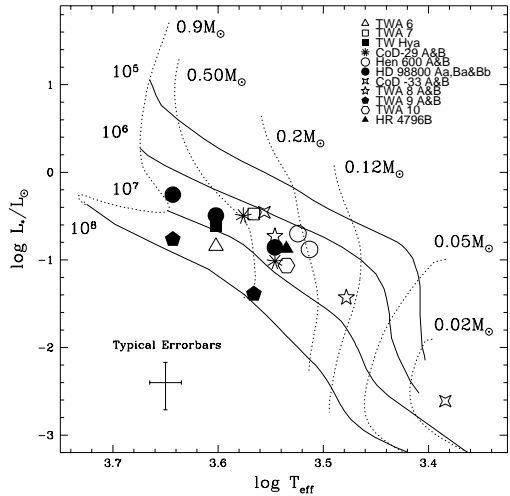}{2.23in}{0}{35}{35}{-110}{-60}
\caption{\scriptsize The TW Hya Association stars plotted on an HR
diagram along with the PMS tracks of D'Antona \& Mazzitelli (1997 -
with 1998 update).  All stars are assumed to be at 50 pc, except those
with measured Hipparcos parallaxes which appear as filled symbols (TW
Hya, HD 98800, HR 4796B, TWA 9 A\&B).  The vertical error bars
represent a conservative uncertainty of $\pm$15 pc in distance (given
that the 4 TWA members with Hipparcos distances lie between 47 and 67
pc from Earth).  Horizontal bars represent the typical classification
uncertainty of 1 spectral subclass or about $\pm$ 150 K.  HD 98800A is
a known single-line spectroscopic binary and the fainter companion
should have little effect on the derived luminosity and temperature.
Preliminary results from HIRES spectra indicate that Hen 600A, TWA5A
\& TWA 6 may be spectroscopic binaries (Webb, Reid \& Zuckerman 1998).
The indicated positions in Fig. 3 represent the combined system
parameters; the primary effect of an unresolved companion is a slight
increase in observed luminosity.  Placement of TWA 2B is based purely
on $\Delta$K magnitude and the color - spectral type relations of
Kirkpatrick et al. (1994).  TWA 5B's temperature is from its I-K color
and its luminosity from its K magnitude and the bolometric correction
(BC$_{\rm J}$~= 2.05 extrapolated to K) adopted by Luhman, Liebert \&
Rieke (1997).  Due to contamination from the primary and the
uncertainty in the temperature scale, the representative error bars
should be approximately doubled in size for TWA 5B.  Other bolometric
luminosities are based on measured or published J magnitudes (the band
that is least likely to be affected by non-photospheric excesses) and
the bolometric corrections compiled in Hartigan et al. (1994).  The
temperature scale for M dwarfs is highly uncertain.  Here we use are
the spectral types derived in this paper and the temperature scale of
Luhman \& Rieke (1998).}
\end{figure}

        In Fig. 3, we use the photometry obtained at the IRTF
(assuming a distance of 50 pc for those stars without measured
parallax) and the spectral types derived from the LRIS spectra to
place the TWA members on the pre-main sequence (PMS) tracks of
D'Antona \& Mazzitelli (1997).  All the stars, with the possible
exception of TWA 9, are clearly above the zero-age main sequence and,
within the uncertainties, are consistent with a cluster age of $\sim$
10$^7$ years.  TWA 9 A \& B are likely younger than they appear on
Fig. 3 ($\sim$70 Myr).  With EWs of $\sim$ 500 m\AA, their Li lines
are much stronger than any similar stars in the Pleiades
(e.g. Mart\'{i}n 1997).  Possible explanations for their apparent
greater ages include: derived spectral types that are too hot, larger
than expected errors in the photometry, circumstellar extinction or an
error in the Hipparcos distance (the $\sim$20 pc required to put TWA 9
on the 10$^7$ yr isochrone is a 3$\sigma$ deviation).

    The age of the TWA is moderately well constrained.  Detailed study
of HD 98800 yields 10 $^{+10}_{-5}$ Myr (Soderblom et al. 1998) and of
HR 4796 gives 8-10 Myr(Stauffer et al 1995, Jura et al 1998).  Our
data are consistent with these estimates: The TWA stars appear nearly
along the same isochrone ($\sim$ 8 Myr; Fig. 3), though the distance
uncertainty adds substantial scatter.  The high ratio of X-ray to
total luminosity is characteristic of stars evolved beyond the early T
Tauri phase suggesting an age of 10-100 Myr.  And the spectral
characteristics indicate that, within the errors of our measurements
and the classification scheme, all members are likely weak-lined or
classical T Tauri stars.  In our sample of $\sim$ 45 other X-ray
bright stars in this region, {\it none} displayed significant Li I
absorption which would fill in the ``post-T Tauri star gap'' noted by,
e.g., Mart\'{i}n (1997).  This implies that the TWA stars are not the
young tail of a large population of post - T Tauri stars, but rather a
distinct group of old T Tauri stars with an age less than a few times
10$^7$ years.


       A probable companion to TWA 5 was found $\sim$ 2\arcsec ~north
of the primary in shift-and-add processing of the IRTF speckle data
(Fig. 4).  The primary and companion differ in magnitude by $\Delta$K
= 4.7 $\pm$ 0.1.  The companion was also revealed in our LRIS I-band
images following a deconvolution of the primary's point spread
function with the IRAF DAOPHOT package.  Scaling a PSF from a field
star and fitting the wings of the (saturated) primary indicates
$\Delta$I = 6.9 $\pm$ 0.4 magnitudes.  The measured $\Delta$I,
$\Delta$K, and I-K color (Table 1) suggest a spectral type of M8 to
M8.5, according to the color relations of Kirkpatrick \& McCarthy
(1994).

        The companion is shown as a 4-pt. star in Fig. 3.  The tracks
indicate a mass of about 20 M$_{\rm{Jup}}$; this estimate is largely
insensitive to variations in luminosity (i.e. distance and/or age).
The major uncertainty is the photometric error associated with the
I-band deconvolution and the temperature scale for young, late-M
stars.  However, the steepness of I-K color for late-M stars
constrains the uncertainty to within $\sim$1.5 subclasses.
Uncertainties in the spectral-type to temperature conversion are
discussed in Luhman, Liebert \& Reike (1997) and Lowrance et
al. (1998).  Although there is no proper motion confirmation of
companionship, the small separation at relatively high galactic
latitude and the fact that TWA 5B appears to be coeval with the
primary imply that it is likely a physical companion (see also
Lowrance et al. 1998).

\begin{figure}[t]
\figurenum{4}
\epsscale{1.0}
\plottwo{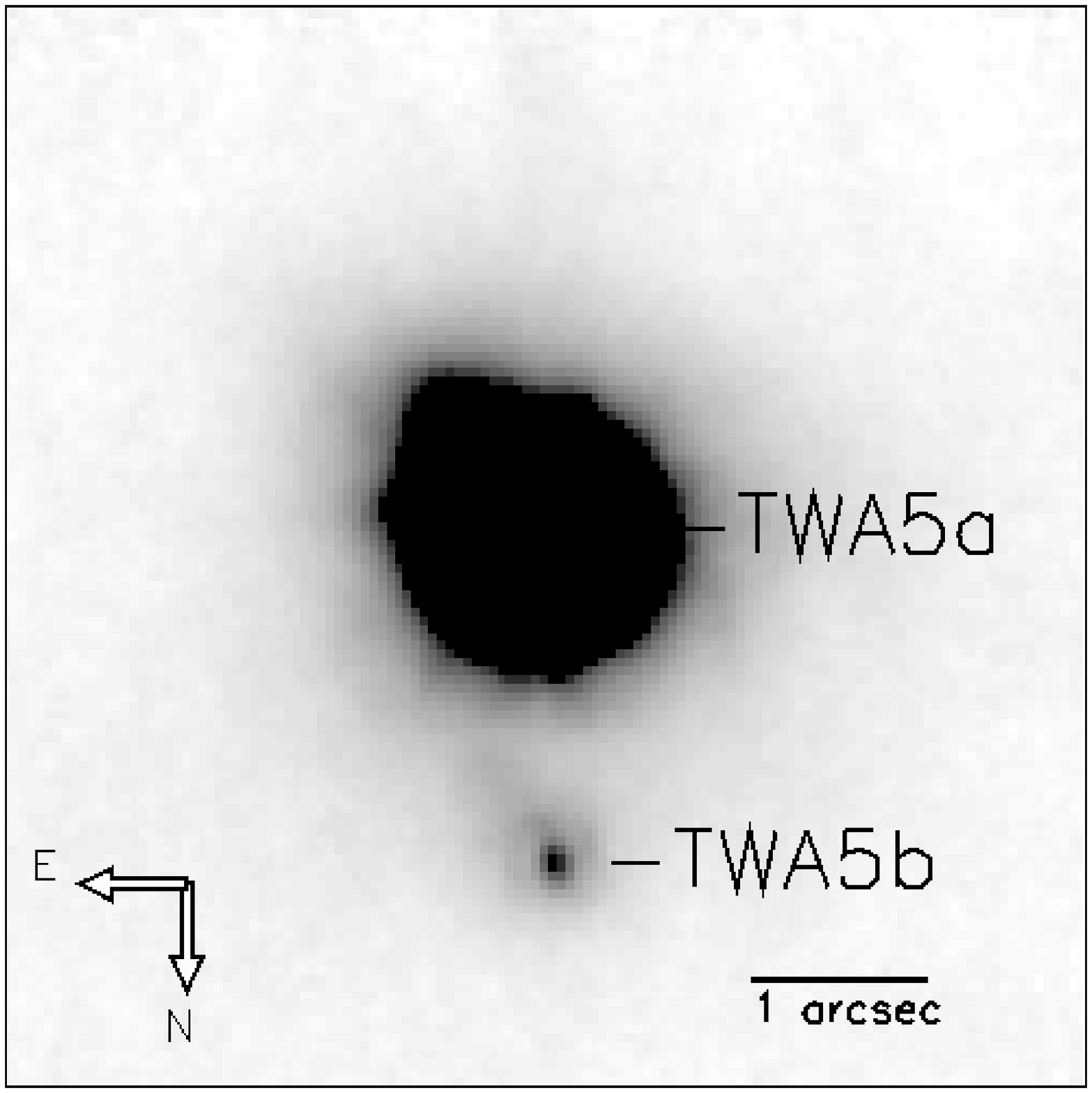}{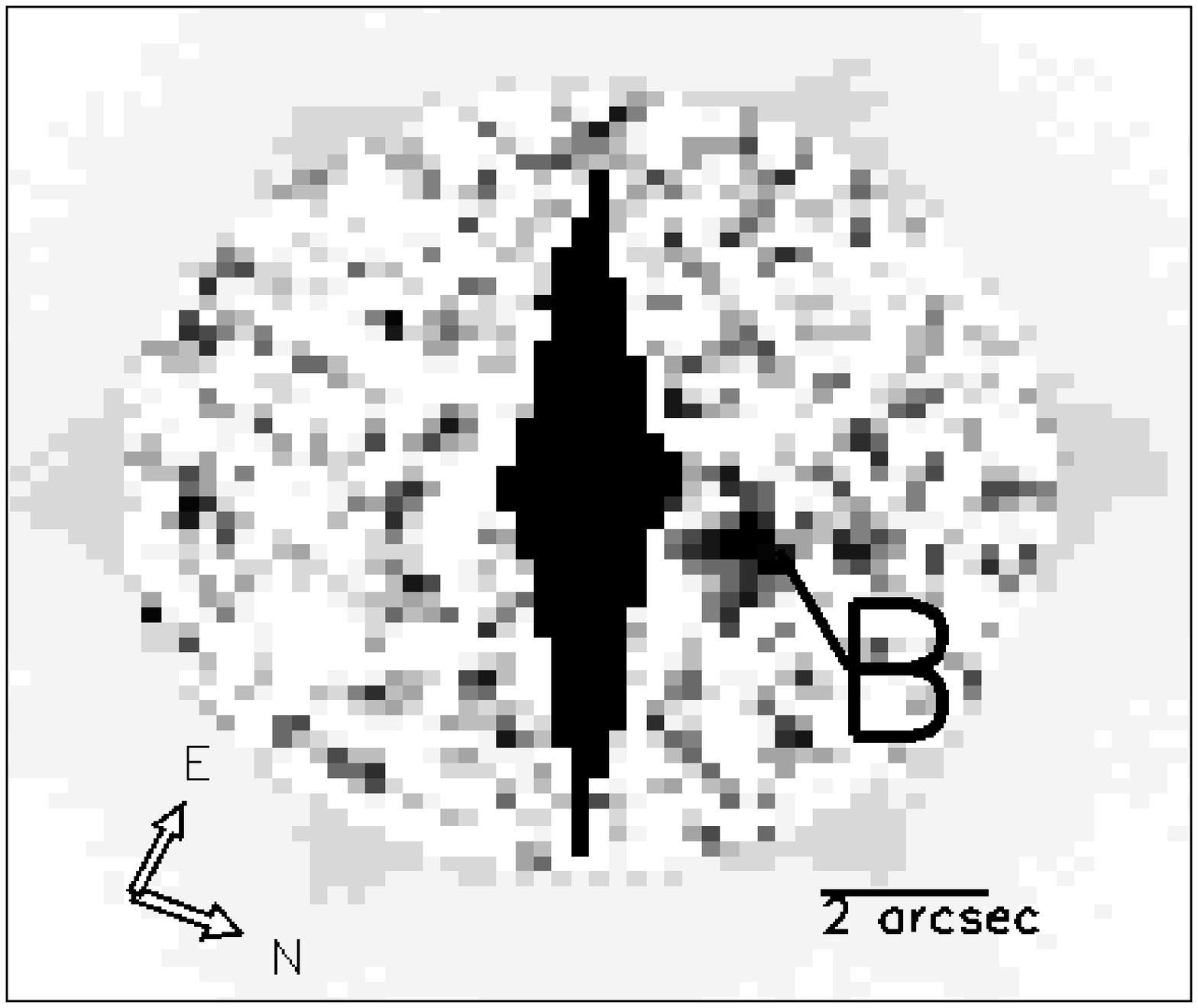}
\caption{\scriptsize a.) A shift-and-add IRTF K-band image showing
TWA 5a (K= 6.83) and TWA 5b (K = 11.5).  The pixel scale is 0\farcs
053 pixel$^{-1}$ and the measured separation is X = +1.77 pixels
(0\farcs 1) and Y = -35.67 pixels (-1\farcs 9).  b.) An LRIS I-band
image of TWA 5.  A gaussian fit to the wings of the saturated primary
has been subtracted out to a radius of 5\arcsec~revealing the B
component.  Saturated pixels and bleeding to adjacent pixels have not
been removed.  The scale is 0\farcs22 pixel$^{-1}$.}
\end{figure}

        Given the rarity of young isolated stars in other parts of the
sky, we believe that HR 4796 A \& B are TWA members despite their
position on the edge of the cluster both in direction
($\sim$18$^{\circ}$ from the cluster center) and in distance (67 pc;
$\sim$15 pc further than the mean distance).  The X-ray emission from
HR 4796, which is emitted by the ``B'' component (Jura et al. 1998),
is consistent with that of the other members.  Thus in every respect
-- age, Li abundance, kinematics \& X-ray brightness -- HR 4796B
appears very similar to the other TWA stars.

\section{Summary}

        Eleven young star systems, totaling at least 19 stars and a
brown dwarf, are now known in the vicinity of the isolated, classical
T Tauri star, TW Hya.  Given the youth of each of the members, the
far-infrared excess emission at four of the TWA systems,
and the common motions demonstrated in Webb, Reid \& Zuckerman (1998),
it is very unlikely that this group of young stars is anything but a
bona fide association of T Tauri stars with a common origin.  This
uniquely close group of young stars is an ideal laboratory to study
star and planetary formation processes.

\acknowledgments

Primary funding for this research is provided by National Science
Foundation (NSF) grant \#9417156 to UCLA.  The SPM program is
supported by grants from the NSF.  IRTF data was funded through NASA
grant \# NAGW-4770 under the Origins of Solar Systems program.  This
research has made use of the Simbad database operated at CDS,
Strasbourg, France.  Most of the data presented herein were obtained
at the W.M. Keck Observatory, made possible by the generous financial
support of the W.M. Keck Foundation and operated as a scientific
partnership among the Caltech, the Univ. of California and the NASA.
We thank J. Kastner for helpful discussions, the referee for helpful
comments and J. Lee, T. Girard \& M. Hazen for assistance with plate
archives.


\begin{deluxetable}{llccccccccccc}
 
\scriptsize
\tablewidth{0pt} 

\tablecaption{Properties of the TWA stars \label{Tab1}}
\tablehead{
    \colhead{TWA}   &
    \colhead{Common}   &
    \colhead{RA}     &
    \colhead{Dec}    &
    \colhead{Spec}  &
    \colhead{H$\alpha$}     &
    \colhead{Li6708\AA}    &
    \colhead{J}  &
    \colhead{H}    &
    \colhead{K}  &
    \colhead{PM RA}    &
    \colhead{PM Dec}  &
    \colhead{Log}   
\nl 
    \colhead{\#}   &
    \colhead{name}   &
    \colhead{J2000}     &
    \colhead{J2000}    &
    \colhead{Type}  &
    \colhead{EW-\AA}     &
    \colhead{EW-\AA}    &
    \colhead{}    &
    \colhead{}  &
    \colhead{}    &
    \colhead{mas yr$^{-1}$}  &
    \colhead{mas yr$^{-1}$}    &
    \colhead{L$_x$/L$_{bol}$}
}
    
\tablecolumns{13}  
\startdata
\sidehead{New members}
 
6\tablenotemark{a}   & \nodata              & 10 18 28.8 & -31 50 02 & K7   & -4.65   & 0.56    & 
  8.79                   & 8.17    & 7.97 & -60     & -20     & -2.92  \nl 
 
7                    & \nodata              & 10 42 30.3 & -33 40 17 & M1   & -4.95   & 0.44    & 
  7.78                   & 7.13    & 6.89 & -120    & -27     & -3.24  \nl 
 
8A                   & \nodata              & 11 32 41.5 & -26 51 55 & M2   & -7.34   & 0.53    & 
  8.37                   & 7.72    & 7.44 & \nodata & \nodata & -2.99  \nl 
 
8B                   & \nodata              & 11 32 41.4 & -26 52 08 & M5   & -16.5   & 0.56    & 
  9.91                   & 9.36    & 9.01 & \nodata & \nodata & \nodata \nl 
 
9A\tablenotemark{b}  & CD-36$^{\circ}$7429 & 11 48 24.2 & -37 28 49 & K5   & -2.35   & 0.46    & 
  8.60                   & 7.95    & 7.68 & -54.1   & -20.0   & -3.02   \nl 
 
9B\tablenotemark{b}  & \nodata              & -6\arcsec  & +0\arcsec & M1   & -5.01   & 0.48    & 
  10.06                  & 9.41    & 9.14 & \nodata & \nodata & \nodata \nl 
 
10                   & \nodata              & 12 35 04.3 & -41 36 39 & M2.5 & -8.39   & 0.46    & 
  9.17                   & 8.55    & 8.19 & -67     & -43     & -3.09   \nl

\sidehead{Previously known members}
 
1                    & TW Hya               & 11 01 51.9 & -34 42 17 & K7   & -220    & 0.39    & 
  8.46                   & 7.65    & 7.37 & -66.9   & -12.4   & -2.86   \nl 
 
2A                   & CD-29$^{\circ}$8887 & 11 09 13.9 & -30 01 39 & M0.5 & -1.89   & 0.49    & 
  7.85                   & \nodata & 7.18 & -90     & -20     & -3.35   \nl 
 
2B                   &                      & +0\farcs 3 &+0\farcs 5 & M2   & \nodata & \nodata & 
  9.09                   & \nodata & 7.99 & \nodata & \nodata & \nodata \nl 
 
3A\tablenotemark{a}  & Hen 600A             & 11 10 28.0 & -37 31 53 & M3   & -21.8   & 0.53    & 
  8.22                   & 7.60    & 7.28 & -112    & -11     & -3.30   \nl 
 
3B                   & Hen 600B             & -0\farcs 8 & -1\farcs 2& M3.5 & -7.14   & 0.54   & 
  8.63                   & 8.07    & 7.80 & \nodata & \nodata & \nodata \nl 
 
4                    & HD 98800             & 11 22 05.3 & -24 46 40 & K5   & 0       & 0.36    & 
  6.44                   & 5.82    & 5.65 & -85.5   & -33.4   & -3.44   \nl 
 
5A\tablenotemark{a}  & CD-33$^{\circ}$7795 & 11 31 55.4 & -34 36 27 & M1.5 & -13.4   & 0.57    & 
  7.71                   & 7.06    & 6.83 & -86     & -21     & -2.95   \nl
 
5B                   &                      & +0\farcs 1 &+1\farcs 9 & M8.5 & \nodata & \nodata & 
  I=15.8\tablenotemark{d}& 12.1 & 11.5 & \nodata & \nodata & \nodata \nl
 
11A\tablenotemark{c} & HR 4796A             & 12 36 01.3 & -39 52 09 & A0   & \nodata & \nodata & 
  5.80                   & 5.80    & 5.80 & -67.1   & -55.9   & \nodata \nl
 
11B\tablenotemark{c} & HR 4796B             & -5\farcs 3 &-5\farcs 5 & M2.5 & -3.5    & 0.55    & 
  9.32                   & 8.57    & 8.36 & \nodata & \nodata & -3.30\tablenotemark{e} \nl 
 
\enddata

\tablenotetext{a}{TWA 6, TWA 5A and Hen 600A are possible spectroscopic
binaries (Webb, Reid \& Zuckerman 1998).  Parameters listed are
the total of the unresolved systems.}
 
\tablenotetext{b}{TWA 9 was independently discovered by Jensen, Cohen
\& Neuh\"{a}user (1998)}
 
\tablenotetext{c}{HR 4796 was previously known to be a young star
system, but not known as a TWA member.} 
 
\tablenotetext{d}{I-band measurement for TWA 5B is from LRIS imaging
data.  The presence of the bright, nearby primary induces an
uncertainty of $\sim$ $\pm$ 0.4 magnitudes.}
 
\tablenotetext{e}{Pointed ROSAT observations indicate the X-rays are
emitted by the B component of the HR 4796 system Jura et
al. (1998).}

\tablecomments{TWA numbers are ordered by RA, beginning with the first
five known members from Kastner et al. (1997), followed by the new
members presented in this paper.  Positions are Equinox J2000 from the
USNO A1.0 catalog except HR 4796 (from SIMBAD) and close companions
whose separations are computed from our observations.  Spectral types
are assigned utilizing our LRIS spectra and standards from Montes \&
Mart\'{i}n (1998).  Equivalent widths (EW) of H$\alpha$ \& Li were
derived from our LRIS spectra except for HR 4796 A \& B - data from
Stauffer et al. (1995) \& Jura et al. (1998), respectively.  The
listed EW of the Li line represents a gaussian fit to the red side of
the line to avoid contribution from nearby, blended Fe line.  The
uncertainty in the Li EW is typically $\sim$ 50 m\AA.  NIR photometry
is from the IRTF, except TW Hya, HD 98800 and HR 4796 whose values are
from Rucinski \& Krautter (1983), Zuckerman \& Becklin (1993) and Jura
et al. (1993), respectively.  IRTF J \& H uncertainties are typically
$\pm$ 0.1 magnitudes, K-band speckle uncertainties are $\pm$0.07
magnitudes.  Proper motions (PM) are derived from the SPM database,
except those with Hipparcos measurements which are denoted by PM's
quoted to the nearest tenth.  L$_x$ / L$_{bol}$ is derived from the
ROSAT All-Sky Survey and the J band magnitudes.  All components of
multiple systems are assumed to contribute to the X-ray flux, hence
L$_{bol}$ represents the total system flux.  The exceptions are HR
4796 (see note $^e$) and TWA 8 (where the X-rays appear to originate
from the primary). }

\end{deluxetable}

\end{document}